\documentclass{amsart}
\pdfoutput=1 

\usepackage{hyperref}  

\usepackage[backend = biber,
	sorting = none,
    style = authoryear,
    giveninits = true,  
    dashed = false,     
    uniquename = false  
]{biblatex}
\DeclareCiteCommand{\cite}[\mkbibemph]
  {\usebibmacro{prenote}}%
  {\usebibmacro{citeindex}%
   \usebibmacro{cite}}
  {\multicitedelim}
  {\usebibmacro{postnote}}

\DeclareBibliographyCategory{trans}
\DeclareBibliographyCategory{eval}
\DeclareBibliographyCategory{russo}
\DeclareBibliographyCategory{other}
\DeclareBibliographyCategory{myself}

\addtocategory{myself}{Nilsson2024minimax}

\addtocategory{trans}{Rinner1958}
\addtocategory{trans}{Hirvonen1959}
\addtocategory{trans}{Hirvonen1960}
\addtocategory{trans}{Berger1960}
\addtocategory{trans}{Gersten1961}
\addtocategory{trans}{Morrison1961}
\addtocategory{trans}{Purcell1961}
\addtocategory{trans}{Brooks1962}
\addtocategory{trans}{Hirvonen1963}
\addtocategory{trans}{Baird1964}
\addtocategory{trans}{Hirvonen1964}
\addtocategory{trans}{Christ1965}
\addtocategory{trans}{Pick1966}
\addtocategory{trans}{Ecker1967}
\addtocategory{trans}{Sugai1967}
\addtocategory{trans}{Sunkel1967}
\addtocategory{trans}{Mikhailov1967}
\addtocategory{trans}{Pick1967}
\addtocategory{trans}{Heffron1967}
\addtocategory{trans}{Bencini1968}
\addtocategory{trans}{Pavlov1968}
\addtocategory{trans}{Tomelleri1970}
\addtocategory{trans}{Getchell1972}
\addtocategory{trans}{Paul1973}
\addtocategory{trans}{Unguendoli1974}
\addtocategory{trans}{Benning1974}
\addtocategory{trans}{Long1974}
\addtocategory{trans}{Deprit1975}
\addtocategory{trans}{Barteleme1975}
\addtocategory{trans}{Long1975}
\addtocategory{trans}{Hedgley1976}
\addtocategory{trans}{Bowring1976}
\addtocategory{trans}{Vincenty1976}
\addtocategory{trans}{Frohlich1976}
\addtocategory{trans}{Bopp1976}
\addtocategory{trans}{Vincenty1978}
\addtocategory{trans}{Danielsen1978}
\addtocategory{trans}{Penev1978}
\addtocategory{trans}{Pick1978}
\addtocategory{trans}{Vincenty1980}
\addtocategory{trans}{Carlson1980}
\addtocategory{trans}{Chen1981}
\addtocategory{trans}{Brookshire1981}
\addtocategory{trans}{Heikkinen1982}
\addtocategory{trans}{Terrence1982}
\addtocategory{trans}{Sampson1982}
\addtocategory{trans}{Scholdbauer1984}
\addtocategory{trans}{Bowring1985}
\addtocategory{trans}{Lupash1985}
\addtocategory{trans}{Ozone1985}
\addtocategory{trans}{Eissfeller1985}
\addtocategory{trans}{Sofair1985}
\addtocategory{trans}{DelFrate1985}
\addtocategory{trans}{Pick1985}
\addtocategory{trans}{Wei1986}
\addtocategory{trans}{Benning1987}
\addtocategory{trans}{Borkowski1987}
\addtocategory{trans}{Heck1987}
\addtocategory{trans}{Romao1987}
\addtocategory{trans}{Borkowski1988}
\addtocategory{trans}{Nautiyal1988}
\addtocategory{trans}{Ohlson1988}
\addtocategory{trans}{Czarnecki1988}
\addtocategory{trans}{Levin1988}
\addtocategory{trans}{Borkowski1989}
\addtocategory{trans}{Burchfield1990}
\addtocategory{trans}{Lapain1991}
\addtocategory{trans}{Grafarend1991}
\addtocategory{trans}{Stoppini1992}
\addtocategory{trans}{Hsu1992}
\addtocategory{trans}{Lin1992}
\addtocategory{trans}{Wise1992}
\addtocategory{trans}{Croceto1993}
\addtocategory{trans}{Zhu1993}
\addtocategory{trans}{Lin1993}
\addtocategory{trans}{Zhu1994}
\addtocategory{trans}{Lin1995}
\addtocategory{trans}{Sofair1995}
\addtocategory{trans}{Grafarend1995}
\addtocategory{trans}{Hekimoglu1995}
\addtocategory{trans}{Toms1995}
\addtocategory{trans}{Toms1996}
\addtocategory{trans}{Ohlson1996}
\addtocategory{trans}{Sudano1997}
\addtocategory{trans}{Heindl1997}
\addtocategory{trans}{Sofair1997}
\addtocategory{trans}{Toms1998}
\addtocategory{trans}{Toms1998b}
\addtocategory{trans}{Fotiou1998}
\addtocategory{trans}{Fukushima1999}
\addtocategory{trans}{Sjoberg1999}
\addtocategory{trans}{Sofair2000}
\addtocategory{trans}{Grafarend2000}
\addtocategory{trans}{Nievergelt2000}
\addtocategory{trans}{You2000}
\addtocategory{trans}{Graferend2001}
\addtocategory{trans}{Guo2001}
\addtocategory{trans}{Pollard2002}
\addtocategory{trans}{Jones2002}
\addtocategory{trans}{Vermeille2002}
\addtocategory{trans}{Wu2003}
\addtocategory{trans}{Vermeille2004}
\addtocategory{trans}{Pollard2005}
\addtocategory{trans}{Zhang2005}
\addtocategory{trans}{Awange2005}
\addtocategory{trans}{Awange2005b}
\addtocategory{trans}{Fukushima2006}
\addtocategory{trans}{Gade2007}
\addtocategory{trans}{Featherstone2008}
\addtocategory{trans}{Sjoberg2008}
\addtocategory{trans}{Feltens2008}
\addtocategory{trans}{Shu2008}
\addtocategory{trans}{Turner2009}
\addtocategory{trans}{Gade2010}
\addtocategory{trans}{Shu2010}
\addtocategory{trans}{Li2010}
\addtocategory{trans}{Behnabian2010}
\addtocategory{trans}{Karney2011}
\addtocategory{trans}{Ligas2011}
\addtocategory{trans}{Turner2011}
\addtocategory{trans}{Vermeille2011}
\addtocategory{trans}{Civicioglu2012}
\addtocategory{trans}{daSilva2012}
\addtocategory{trans}{Turner2013}
\addtocategory{trans}{Ligas2013}
\addtocategory{trans}{Zeng2013}
\addtocategory{trans}{Eberly2013}
\addtocategory{trans}{Osen2017}
\addtocategory{trans}{Turner2016}
\addtocategory{trans}{Hmam2018}
\addtocategory{trans}{Uteshev2018}
\addtocategory{trans}{Panou2018}
\addtocategory{trans}{Medvedev2018}
\addtocategory{trans}{Tatar2018}
\addtocategory{trans}{Claessens2019}
\addtocategory{trans}{Medvedev2019}
\addtocategory{trans}{Panou2019}
\addtocategory{trans}{Kadaj2020}
\addtocategory{trans}{Eleiche2020}
\addtocategory{trans}{Medvedev2020}
\addtocategory{trans}{Knopp2021}
\addtocategory{trans}{Panou2021}
\addtocategory{trans}{Guo2023}
\addtocategory{trans}{Quan2024}
\addtocategory{trans}{Eleiche2024}
\addtocategory{trans}{Blewitt2024}
\addtocategory{eval}{Voll1990}
\addtocategory{eval}{Barbarella1993}
\addtocategory{eval}{Barrio1993}
\addtocategory{eval}{Laskowski1991}
\addtocategory{eval}{Keeler1998}
\addtocategory{eval}{Gerdan1999}
\addtocategory{eval}{Seemkooei2002}
\addtocategory{eval}{Fok2003}
\addtocategory{eval}{Burtch2006}
\addtocategory{eval}{Wahlberg2009}
\addtocategory{eval}{Vega2009}
\addtocategory{eval}{Yildirim2011}
\addtocategory{eval}{Lee2014}
\addtocategory{eval}{Bajorek2014}
\addtocategory{eval}{Kumi2016}
\addtocategory{eval}{Lopes2018}
\addtocategory{eval}{Ward2020}
\addtocategory{eval}{Vita2020}
\addtocategory{eval}{Eleiche2022}
\addtocategory{other}{Dorrie1948}
\addtocategory{other}{Heiskanen1967}
\addtocategory{other}{Mayer1978}
\addtocategory{other}{Torge1975}
\addtocategory{other}{Vanicek1982}
\addtocategory{other}{Rapp1984}
\addtocategory{other}{Churchyard1985}
\addtocategory{other}{Vincenty1985}
\addtocategory{other}{HofmannWellenhof1997}
\addtocategory{other}{Claessens2007}
\addtocategory{other}{Zanevicius2010}
\addtocategory{other}{Awange2010}
\addtocategory{other}{Alberto2019}
\addtocategory{other}{Mazzucato2021}
\addtocategory{russo}{Butkevich1962}
\addtocategory{russo}{Laping1962}
\addtocategory{russo}{Butkevich1963}
\addtocategory{russo}{Laping1964}
\addtocategory{russo}{Andreev1966}
\addtocategory{russo}{Mikhailov1966}
\addtocategory{russo}{Butkevich1967}
\addtocategory{russo}{Pavlov1967}
\addtocategory{russo}{Butkevich1968}
\addtocategory{russo}{Daskalova1969}
\addtocategory{russo}{Izotov1969}
\addtocategory{russo}{Pejchev1969}
\addtocategory{russo}{Pavlov1970}
\addtocategory{russo}{Polevoj1970}
\addtocategory{russo}{Ganshin1970}
\addtocategory{russo}{Pavlov1971}
\addtocategory{russo}{Pavlov1971b}
\addtocategory{russo}{Zlatanov1972}
\addtocategory{russo}{Pavlov1973}
\addtocategory{russo}{Czarnecki1972}
\addtocategory{russo}{Daskalova1973}
\addtocategory{russo}{Penev1973}
\addtocategory{russo}{Morozov1974}
\addtocategory{russo}{Anufriev1974}
\addtocategory{russo}{Pavlov1975}
\addtocategory{russo}{Andreev1975}
\addtocategory{russo}{Morozov1978}
\addtocategory{russo}{Chen1979}
\addtocategory{russo}{Penev1980}
\addtocategory{russo}{Zeng1981}
\addtocategory{russo}{Butkevich1982}
\addtocategory{russo}{Butkevich1985}
\addtocategory{russo}{Aleksic1987}
\addtocategory{russo}{Balandin1988}
\addtocategory{russo}{Medvedev1994}
\addtocategory{russo}{Medvedev1995}
\addtocategory{russo}{Medvedev2000}
\addtocategory{russo}{Baldini2006}
\addtocategory{russo}{Shu2009}
\addtocategory{russo}{Poleshenkov2011}
\addtocategory{russo}{Ogorodova2011}
\addtocategory{russo}{Balandin2012}
\addtocategory{russo}{Penev2012}
\addtocategory{russo}{Efanov2013}
\addtocategory{russo}{Ogorodova2014}
\addtocategory{russo}{Guo2014}
\addtocategory{russo}{Medvedev2014}
\addtocategory{russo}{Kravchenko2015}
\addtocategory{russo}{Medvedev2016}
\addtocategory{russo}{Medvedev2016b}
\addtocategory{russo}{Ogorodova2017}
\addtocategory{russo}{Kureniov2017}
\addtocategory{russo}{Penev2021}
\addtocategory{russo}{Afonin2021}

\title[A comprehensive reference listing]{A comprehensive Cartesian to geodetic coordinate transformation reference listing}
\author{John-Olof Nilsson\\{\scriptsize \lowercase{jnil02@kth.se}}}
\date{}

\nocite{*}

\begin{document}

\begin{abstract}
A comprehensive listing of over 230 references related to Cartesian to geodetic coordinate transformations are provided. The references are coarsely sorted and DOI:s or URL:s are provided to the extent possible to facilitate studies of the material.
\end{abstract}

\maketitle

\begingroup
  \boolfalse{citerequest} 

\section{Introduction}
Computing geodetic coordinates $\{\phi,\lambda,h\}$ from Cartesian $\{x,y,z\}$ coordinates
, i.e. inverting
\begin{equation*}
\begin{split}
x &= (N+h)\cos(\phi)\cos(\lambda)\\
y &= (N+h)\cos(\phi)\sin(\lambda)\\
z &= ((N(1-e^2)+h)\sin(\phi)
\end{split}
\end{equation*}
where $N=(1-e^2\sin(\phi))^{-1/2}$,
is a problem with surprising depth. Unfortunately, to date, there is no comprehensive review of the work on the problem. In combination with the high number of publications, the number of languages and journals, including old non-digitalized/non-searchable ones, in which results have been published and the half-century long time span of the work, this makes assessing novelty of new work or reading up on existing work challenging. To partially remedy this problem, here a comprehensive listing of over 230 related references, spanning more than 6 decades, are provided. To the extent possible, DOI:s or URL:s are provided to facilitate accessing the references. This listing significantly expands on earlier listings, e.g.~\cite{Lapain1991,Featherstone2008,Mazzucato2021}, and constitute a different but important contribution to the field. It is my hope that this collection of common and rare references will reduce wasted time and research effort and contribute to better citing of previous work within the field.
This listing was prepared to assess the novelty of and as supplementary material for
\vspace{4mm}

\printbibliography[category={myself},heading=none]

\vspace{4mm}

\noindent
The publication also contains some further history on the subject and comments about the older references.


Publications are coarsely group in \emph{Algorithms}, \emph{Benchmarks}, \emph{Soviet Bloc and Chinese Literature} and \emph{Other}. Publications within each group are listed in oldest-first chronological order.  Publications which have not been accessed are labeled accordingly and, in most cases, accompanying citations of one ore more citing sources are provided. Note, work on non-altitude dependent transformations and transformations for triaxial ellipsoids are not included in this listing.
%


\section{Algorithms}
The following publications describes algorithms for carrying out the Cartesian to geodetic coordinate transformation.
\vspace{5mm}

\printbibliography[category={trans},heading=none]

\section{Benchmarks}
The following publications contains reviews and benchmarks of previously published algorithms. Note, none of the publications contains any comprehensive review of the field and most publications only benchmark a small set of algorithms.
\vspace{5mm}

\printbibliography[category={eval},heading=none]

\section{Soviet Bloc and Chinese Literature}
The following literature constitutes a largely hard-to-access literature in primarily Russian and Chinese. The references primarily comes from~\cite{Lapain1991}, a few authors who have published both in the Soviet Bloc journals and in western journals and the journal \emph{Geodesy and cartography = Geodezia i Kartografia} which have an digital archive with references and refbacks. However, many sources are probably missed since not all titles are translated.
\vspace{5mm}

\printbibliography[category={russo},heading=none]

\section{Other}
The following publications are frequently cited books or publications containing material related to the coordinate transformations.
\vspace{5mm}

\printbibliography[category={other},heading=none]





\endgroup 

\end{document}